\documentstyle[aps]{revtex}

\begin{document}

\title{The U(1) and Strong CP Problem From the Loop Formulation
Perspective}

\author{Hugo Fort and R.  Gambini\footnote{Associate member of
ICTP} \\ Instituto de F\'{\i}sica, Facultad
de Ciencias, \\ Tristan Narvaja 1674, 11200 Montevideo, Uruguay}

\maketitle

\begin{abstract}
We show how the gauge invariant formulation of QCD in terms 
of loops is free from a hidden $\theta$ parameter and offers an
alternative way to solve the $U(1)_A$ 
puzzle. 
\end{abstract}

\vspace{2mm}

Quantum Chromodynamics (QCD) in its standard formulation has a
non-trivial topological structure which is manifest through the
existence of {\em large} gauge transformations -- characterized by a
topological integer or {\em winding number} $n$ -- in addition to the
ordinary {\em small} gauge transformations
(generated by the Gauss's law) with $n=0$.  This implies
that there is an infinite set of degenerate vacuum states, each labeled
by its $n$.  Instanton solutions provide a mechanism of ``vacuum
tunneling" between topological inequivalent $n$-vacua \cite{tH76}.  So
that the ``true" vacuum -- the so-called $\theta$ vacuum -- is a linear
superposition of $n$-vacua \cite{cdg76} and as a consequence the theory
possesses a hidden parameter, the vacuum angle $\theta$.

This non-trivial topological structure of 
QCD vacuum has two remarkable effects.  The
first effect that was soon pointed out \cite{tH76} is that it 
offers a solution of the ``$U(1)_A$ problem".  
In a nutshell the $U(1)_A$ puzzle is:  
the approximate axial symmetry is known to be broken, so {\em where
is the corresponding quasi-Nambu-Goldstone boson?}. The answer to this
question is a little bit involved and can be traced to the fact
that $U(1)_A$ is broken, by virtue of the axial
anomaly, to the discrete symmetry $Z_{2N_f}$, where $N_f$ is the 
number of flavors.  
This incomplete breaking of the axial symmetry opens
the possibility that this breaking is not accompanied by a
Nambu-Goldstone boson \cite{mar}.
This solution does not depend on
the value of the $\theta$ parameter associated with the non-perturbative
QCD vacuum.  The second effect is that a non-zero value of $\theta$
implies a violation of CP invariance in strong interactions.  This can
be understood as follows:  the rich structure of Yang-Mills vacuum
corresponding to tunnelling between states of different winding number
gives rise to an effective Lagrangian term proportional to $\theta$
times the Chern-Pontryagin density $F \wedge F$, which violates P and CP
conservation.  
Strong interactions processes conserve the CP symmetry, 
therefore the ``strong CP" problem is:  {\em why is $\theta=0$?}.  
The same
$\theta$ vacuum that solves the $U(1)_A$ problem in QCD creates the
strong CP problem.  It is important to note that the existence of the
$\theta$ vacua and the value of the $\theta$ parameter are different
questions.  Different solutions to avoid the CP-problem have been
considered but all present some drawback. 
A first alternative is to solve the problem by postulating that one 
of the quark masses (presumably $m_u$) is equal to zero.  
However, a massless quark contradicts the current algebra 
calculations of the quark masses \cite{le}.
A second proposal, and by far the most popular approach, 
is the Peccei-Quinn (PQ) mechanism
\cite{pq77} which introduces a new additional chiral U(1) symmetry which
allows to rotate the $\theta$ parameter to zero.  Unfortunately, a
by-product of this mechanism is the generation of a Nambu-Goldstone
boson, the {\em axion}, which has eluded detection.
A third proposed solution is simply to set $\theta=0$ based on
mathematical grounds \cite{kugo-ojima}.  
However, this proposal relies on a formulation of
QCD which is still not completely settled \cite{mar}.

Here we propose an entirely different point of view on the  
nature of the $U(1)_A$ anomaly and the CP problem which emerges when one
considers the gauge invariant formulation of Yang-Mills theories 
provided by the {\em loop representation} \cite{gt}.    
Our starting point is the fact that the standard definition of QCD 
is in terms of local fields, quarks and gluons,
but the physical excitations are extended composites:  mesons and
baryons. The loop representation is an alternative
quantum formulation of Yang-Mills theories directly in terms of
paths or strings associated to these extended excitations.
The basis of this formulation can be traced to the 
idea of describing gauge theories
explicitly in terms of Wilson loops or holonomies \cite{p},\cite{mami}
since Yang \cite{y} noticed their important role for a complete
description of gauge theories.  
The loops replace the
information furnished by the vector potential (the connection)
providing the necessary and sufficient observable information
about gauge theories. A
description in terms of loops or strings, besides the general advantage
of only involving the gauge invariant quantities, is 
appealing because all the gauge invariant operators have a 
simple geometrical meaning when realized in the loop space.  
Moreover, the
interest on loops relies on the fact that it was realized that this
formalism goes beyond a simple gauge invariant description and in fact
it provides a natural geometrical framework to treat gauge theories and
quantum gravity.  The introduction by Ashtekar \cite{a} of a new set of
variables that cast general relativity in the same language as gauge
theories allowed to apply loop techniques \cite{rs} as a natural
non-perturbative description of Einstein's theory \cite{gapu}. 
An interesting question to address is:  
do loops provide a different physical picture of gauge theories
to the one which emerges from the conventional formulation in
terms of fields?  Or on the contrary, they only provide a particular
and more economic mathematical description?  In this letter we discuss
this issue and in particular we focus on the first option. 
However, if one supports the
second option, it is also  possible to ``enlarge'' the loop Hilbert 
space in such a way to reproduce exactly the ordinary results.  
We will show that several recent calculations involving the
inclusion of fermions in the loop representation \cite{gmuv}-\cite{f97}
provide a different picture of the two previously mentioned problems:
{\bf i)} there is no hidden $\theta_{QCD}$ parameter coming from the 
strong interaction sector, and 
{\bf ii)} the absence of a Goldstone boson for the broken U(1)$_A$ 
is straightforward without 
resorting to the instanton-based mechanism.

The first thing to note is that, obviously, 
{\em in the loop representation the
distinction between large and small gauge transformations is
meaningless:  the states are invariant under both}. In fact, 
the states may be considered as linear combinations of
Wilson loops, and consequently, due to the cyclic property of the trace,
they are invariant under small and large gauge transformations.
Therefore, the generator of large gauge transformations is trivially
equal to one, and the vacuum degeneration is absent. 
Thus {\em the vacuum is unique and no
$\theta_{QCD}$ parameter is hidden}.
As we shall discuss in more detail later
on, the Hilbert space in the loop representation is isomorphic to a 
subspace of the ordinary space.  The solution to the $U(1)_A$ problem 
in this description is quite different and do not requires to resort to
topological instantons. In order to explain this point, 
let us recall how fermions are included in the loop 
representation giving rise to the open path representation 
or {\em P-representation} \cite{fg}.  To introduce
a gauge invariant basis one starts by considering an (overcomplete) set
of commuting gauge invariant operators of the form 
$\psi_d^\dagger(x) H(P_x^y)
\psi_u (y)$, where $\psi_d$ and $\psi_u$ respectively are the up 
and down components of the spinor and $H(P_x^y)$ is the holonomy 
associated with the open path $P_x^y$  going from $x$ to $y$.
In a lattice with staggered or Susskind's fermions \cite{sus} i.e.
with a single component fermion field $\phi(n)$ defined at each site 
$n$ such that
$$
\psi_u(n)=\phi(n),\;\;\;\;n \mbox{even}\,,
$$
\begin{equation}
\;\;\,\psi_d(n)=\phi(n),\;\;\;\;n \mbox{odd}\,,
\end{equation} 
this definitions turn out to be
equivalent to consider gauge invariant operators starting at even sites
and ending at odd sites.  This set of operators define the open path
basis $|P>$.  
Notice that the choice of this basis 
automatically breaks the remnant discrete chiral invariance
of the usual lattice staggered fermion formulation \cite{sus}. 
Now a lattice translation by odd integers of
a basis vector is not a basis vector.  Thus we see that the anomaly
responsible of this breaking is present in the P-representation
from the very beginning.  This fact cannot be unexpected.  
As it is well known
\cite{Ja}, the anomaly occurs as a consequence of the incompatibility of
two classical symmetries- gauge and chiral invariance- at the quantum
level.  It happens that the gauge symmetry may only be preserved at the
prize of sacrificing the chiral symmetry which become anomalously
broken.  The P-representation deals with gauge invariant quantities
and hence has no chance to implement the chiral symmetry.
The chiral symmetry breaking is also apparent in the 
recently proposed Lagrangian counterpart of the P-representation, 
the {\em worldsheet formulation} with dynamical fermions  
\cite{afg}. The worldsheet partition function of lattice QED  
is expressed as a sum over 
surfaces with border on self-avoiding fermionic loops 
with Boltzmann weights which are not
invariant under odd translations so that the action has no
remnant of the chiral symmetry. 

To illustrate the previous points we shall resort to a simple model:
(1+1) QED with massless fermions or the Schwinger model.
This model is rich enough to share  
with 4-dimensional QCD the issues we are concerned with, namely
the topological structure which gives rise to the $\theta$ vacuum
and the breaking of the chiral symmetry with an axial anomaly.
The Schwinger model in the P-representation has been studied 
both in the continuum \cite{gmuv}
and on the lattice \cite{afa97},\cite{f97}.
The analytical continuum study of ref. \cite{gmuv} 
showed that the divergence of the axial current 
is non zero. Nevertheless, it is not clear if it is possible    
to cast it as the divergence of a Chern-Simons density.
In ref.\cite{afa97} using a Hamiltonian finite lattice analysis a
non-null chiral condensate completely consistent with 
the known value in the continuum
\begin{equation} 
\beta^{\frac{1}{2}} <\bar{\psi} \psi \,> =
\frac{\mbox{e}^\gamma}{2\pi^{3/2}}=0.15995, 
\label{eq:chiral-cont}
\end{equation} 
where $\beta=\frac{1}{e^2}$ (the inverse of the square of
the coupling constant $e$) and $\gamma$ is the Euler constant.
This non-zero value of the chiral condensate is a manifestation of the
axial anomaly. Furthermore, a Monte Carlo simulation of this model 
\cite{f97} shows that the chiral symmetry is
broken for the {\em strictly} massless case and it produces 
a chiral condensate which converges in the weak
coupling (continuum) limit (once more) to its known exact  
value. It is worth mention that this is
a clear difference to what happens in an ordinary simulation in terms of
fields, for which in the massless case, given enough time, the system
rotate through all the degenerate minima so that $<\bar{\psi} \psi \,>
=0$.  Therefore, one has to calculate this order parameter for several
small non-zero masses $m$ to get the sensible limit at 
$m \rightarrow 0$. On the contrary, as in the lattice 
loop representation there
is not a discrete chiral symmetry, remnant from the continuous one, it 
is enough to study the massless case.

All this evidence leads to conclude that there is not $U(1)_A$
problem in the P-representation, in fact 
there is not $U(1)_A$ symmetry at the second quantized level 
from the very beginning.

We would like to mention that, even in the loop representation, it is
always possible to introduce a $\theta$ parameter by hand.
Let us recall that,
from the canonical point of view, in the ordinary representation
the generator of large gauge transformations 
has a non trivial action on the wave functions of the pure gauge 
theory given by \cite{Ja},\cite{Huang} 
\begin{equation}
\Omega_n\Psi[A]= \Psi[A_{g_n}],
\end{equation}
where $A_{g_n}$ is obtained by acting on $A$ with a large gauge
transformation with winding number $n$. The Gauss law does not enforce
the invariance under this type of transformations.
Now, as the Hamiltonian $\hat H$ and the unitary operator $\Omega_n$
commute, they are simultaneously diagonalizable
\begin{equation}
\Omega_n\Psi_\theta[A]=\exp [\, i\theta n ]\,\Psi_\theta[A],
\end{equation}
\begin{equation}
{\hat H}\Psi_\theta[A]= E_\theta\Psi_\theta[A].
\end{equation}
For a fixed value of $\theta$ it is possible to introduce a change of
variable of the corresponding $\Psi_\theta$
\begin{equation}
\Phi_\theta[A]=\exp{-i\theta S_{CS}[A]} \Psi_\theta[A]
\end{equation}
such that $\Phi$ is invariant under both small and large gauge
transformations and satisfies the following eigenvalue equation
\begin{equation}
{\hat H}_\theta\Phi_\theta[A] = E_\theta \Phi_\theta[A]
\end{equation}
Notice that now the Hamiltonian depends on $\theta$. It may be obtained
by following the usual canonical procedure from the action
$S_\theta = S_{YM} + \theta \int F \wedge F$ 
where the second term is the
Chern Pontryagin topological invariant. This term does not modify the
field equations because it only adds a surface contribution.
In the loop representation $\Omega_n$ is a trivial operator, 
proportional to the identity and one only has a description of one 
of the gauge invariant sectors with wavefunctions $\Phi$ characterized 
by a value of $\theta$. While in the standard approach, if we start 
with an action with $\theta=0$ the $\theta$ vacuum, associated to
large transformations, still appears, in the loop description of QCD
there is a CP violation term only if this is introduced by hand. 

To conclude, we want to point out that the lattice is the natural arena 
to discuss the P-representation of 4 dimensional QCD. In fact,
a mathematically rigorous 4 dimensional formulation only exist 
on the lattice
and is in this framework where a comparasion with the standard
approach in terms of fields makes sense. In this 
conventional formulation  there is a twofold degeneracy connected with
chiral symmetry i.e. there are two vacuum states which transform
one into each
other by interchanging odd and even sites \cite{brskjss}. In order to
compute the hadron spectrum the procedure is to modify the Hamiltonian
by adding an {\em irrelevant} term (i.e. an operator which has no
effect in the continuum limit) such that it renders the vacuum 
well determined.
On the other hand, the gauge invariant P-representation selects 
one of the two possible chiral sectors from the very beginning. 
This is consistent to the fact that both sectors are separated
for any value of the mass, in the  continuum limit,  
by an infinite gap, so that the value of the physical observables 
should not be affected.
Indeed this was confirmed for QCD \cite{gs} and for the Schwinger
model \cite{afa97}, \cite{f97}.

{\large \bf Acknowledgements}

\vspace{2mm}

We were supported in part by CONICYT, Projects No. 49 and No. 318.
We are indebted with G. Gonzalez-Sprinberg, R. Mendez and M. 
Reisenberger for valuable discussions.

\end{document}